\documentclass[eng
               ]{jetp}
\twocolumn
\begin{document}
\English
\title{Polarization and speed of gravitational waves in hybrid metric-Palatini f(R)-gravity}

\author{P.~I.}{Dyadina}
\email{guldur.anwo@gmail.com}
\affiliation{Sternberg Astronomical Institute, Lomonosov Moscow State University, Universitetsky Prospekt, 13, Moscow, Russia}

\rtitle{Polarization and speed of gravitational waves in hybrid metric-Palatini f(R)-gravity}
\rauthor{P.~I.~Dyadina}
\abstract
{
The question of the correspondence between the number of degrees of freedom and the number of polarization states in hybrid metric-Palatini f(R)-gravity is considered. It is shown that the scalar transverse breathing and longitudinal modes are a mixture of one polarization state. The speed of gravitational waves propagation in hybrid f(R)-gravity is calculated.
}

\date{15 June, 2022}

\maketitle

\section{Introduction}
The discovery of gravitational-wave radiation has become one of the most important events of recent decades. The existence of gravitational waves (GW) and their direct detection confirmed the consequences of the general relativity (GR) \cite{gw}. Moreover, the data obtained by GW detectors serve to test modified theories of gravity. One way to use GW data for testing the gravitational theories is a comparison of the polarization content predicted by the theory with observational data. The event GW170814 allows to LIGO and VIRGO collaborations to determine that purely tensor polarizations are preferable to scalar and vector ones \cite{gw1}. The data obtained from this event can be used for accurate quantitative test of gravitational theories \cite{liu}. And first of all, it is worth paying attention to the number of polarization modes predicted by the theory.

In GR the gravitational wave has two tensor polarization states, the plus and cross ones. There may exist up to six polarization modes  in alternative metric theories of gravity. For null plane gravitational waves, the six polarizations are classified by the little group E(2) of the Lorentz group developed by Eardley \cite{eardley, eardley1}. According to this classification, the massless scalar field propagates with the transverse breathing mode, and the second scalar mode, the longitudinal one, appeares only in the presence of all the other five modes. However, it was later calculated by Maggiore and Nicolis that in the presence of a massive scalar field, a longitudinal mode appears in addition to the transverse breathing mode, while vector modes are absent \cite{maggiore}. And just recently, the question about the discrepancy between the number of degrees of freedom contained in the theory and the number of polarization modes in the case of massive scalar-tensor theories has been raised \cite{hougong, dong}. This question was considered on the example of Horndeski theory \cite{horndeski}. This model is the most general scalar-tensor theory with second-order field equations. The authors showed that  the polarization state excited by the scalar field form a mix polarization of the
breathing and the longitudinal states and, as a result, it corresponds to one degree of freedom \cite{hougong}.

This work is devoted to the study of the polarization content of hybrid metric-Palatini f(R)-gravity \cite{hybrid1, hybrid2}. The approach of f(R)-gravity allows to extend GR in the simplest way. The f(R)-gravity is based on generalization of the gravitational part of the action as an arbitrary function of the Ricci scalar~\cite{fr1, fr2,fr3}. Such models have become widespread after f(R)-gravity was successfully applied in inflation theories \cite{starobinsky}. The f(R)-gravity is attractive because the accelerated expansion of the Universe is natural consequence of the gravitational theory. In addition, f(R)-gravity is attractive as an alternative to the $\Lambda$CDM model, since it allows one to simultaneously describe inflation in the early Universe and accelerated expansion in the later stages of evolution~\cite{odintsov1, odintsov2, odintsov3, odintsov4, odintsov5, odintsov6, odintsov7, saez}. Also, the f(R)-models are in good agreement with the observational cosmological data and practically do not differ from $\Lambda$CDM~\cite{odintsov8}.

The family of f(R)-theories is divided into two classes: metric one and Palatini one. In the metric approach metric is the only dynamical variable and the action is varied with respect to it only. The Palatini method is based on the idea of considering the connection defining the Riemann curvature tensor to be a priori independent of the metric.
However, some shortcomings of f(R)-theory are manifested both in the metric and in the Palatini approaches. One of the main problems of metric f(R)-gravity is difficulties with passing standard tests in the Solar System \cite{chiba}. Nevertheless, a limited class of viable models in the metric approach exists  \cite{odintsov4, odintsov7, odintsov8}. The feature of such models is chameleon mechanism \cite{khoury, khoury1, hu, odintsov3}. All Palatini f(R)-models lead to microscopic matter instabilities and to unacceptable features in the evolution patterns of cosmological perturbations \cite{koivisto}.

Recently, a theory constructed as a superposition of the Einstein-Hilbert metric Lagrangian with the Palatini f(R)-term has been proposed \cite{hybrid1}. The model was called the hybrid metric-Palatini f(R)-gravity. It unites all positive results of metric and Palatini approaches but lacks their shortcomings. The hybrid f(R)-gravity allows to describe a cosmological large-scale structure, without affecting the Solar System dynamics. An attractive feature of the theory is that it allows non-standard scalar-tensor representation in terms of a dynamical scalar field (unlike the Palatini models), which does not have to be very massive to be consistent with the data obtained from laboratory experiments and Solar System observations. In this theory the scalar field can play an active role in cosmology, without entering the conflict with local experiments \cite{hybrid3}.

The hybrid f(R)-gravity was studied on astrophysical scales from stars to galaxy clusters. It was shown that the virial mass discrepancy in clusters of galaxies can be explained via the geometric terms appearing in the generalized virial theorem \cite{hybrid4}. The hybrid f(R)-gravity also allows to explain the rotational velocities of test particles gravitating around galaxies \cite{hybrid5}. This approach allows to avoid introducing of a huge amount of dark matter. Besides, wormhole \cite{hybrid6} and black hole solutions \cite{bronnikov} were derived, and physical properties of neutron, Bose-Einstein Condensate and quark stars were considered \cite{danila}. In addition, hybrid f(R)-gravity was tested in the Solar System using parametrized post-Newtonian formalism (PPN). The analytical expressions for $\gamma$ and $\beta$ parameters were obtained, and it was proved that other 8 PPN parameters identically equal to zero. It was shown that the light scalar field in hybrid f(R)-gravity does not contradict the experimental data based on all set of PPN parameters \cite{meppn, meshapiro}. Moreover, hybrid f(R)-gravity was tested on the binary pulsars observational data. In addition, the change of the orbital period due to gravitational radiation was obtained in the quasicircular case. Also, for the first time the restriction on the scalar field mass in hybrid f(R)-gravity was found \cite{memnras}. After that, the previous results were generalized to the case of an orbit with eccentricity. The hybrid f(R)-theory was tested using a parameterized post-Keplerian formalism. As a result, it was shown that hybrid f(R)-gravity predicts a wider range of possible masses for the components of binary pulsars in comparison with GR \cite{meppk}.

The hybrid f(R)-gravity can be represented as a scalar-tensor theory with a massive scalar field. Previously, the polarization content has already been studied in the framework of hybrid f(R)-gravity \cite{polarization}. It has been found that the theory contains four polarization states: plus, cross, transverse breathing and longitudinal ones. However, earlier it was shown that the theory has only three degrees of freedom \cite{koivisto1}. Thus, the contradiction that arose earlier in Horndeski theory, exist also in hybrid f(R)-gravity. The purpose of this work is to resolve this inconsistency.

Another effective way to test theories of gravity is to determine the propagation speed of the GW signal predicted by the theory. The kilonova, event GW170817/SHB170817A \cite{gw2}, allows to obtain a limit on the propagation velocity of gravitational waves, which led to the reduction or closure of a large number of modified theories of gravity \cite{dead}. This test was also applied to massive scalar-tensor theories. It was shown that only the Brans-Dicke theory with a massive scalar field completely passes this test. In this work, we also calculate the propagation velocity of gravitational waves in hybrid f(R)-gravity and compare obtained result with observational data.

The structure of the paper is the following. In section~\ref{sec2} we consider an action and  field equations of the hybrid metric-Palatini theory in a general form and in a scalar-tensor representation.  In section~\ref{sec3}, we calculate polarization content of hybrid f(R)-gravity. In section~\ref{sec4} we test hybrid f(R)-gravity using the speed of gravitational waves. We conclude in section~\ref{sec:conclusions} with a summary and discussion.
	
		Throughout this paper the Greek indices $(\mu, \nu,...)$ run over $0, 1, 2, 3$ and the signature is  $(-,+,+,+)$. We will be working in units $h = c = k_B = 1$ throughout the paper. The Jordan frame is used.
		
		\section{Hybrid f(R)-gravity}\label{sec2}
	
	The action of hybrid f(R)-gravity has the form~\cite{hybrid1,hybrid3}:
	\begin{equation}\label{act}
	S=\frac{1}{2k^2}\int d^4x\sqrt{-g}\left[R+f(\mathfrak{R})\right]+S_m,
	\end{equation}
	where $k^2=8\pi G$, $G$ is the Newtonian gravitational constant, $R$ and $\mathfrak{R}=g^{\mu\nu}\mathfrak{R}_{\mu\nu}$ are the metric and Palatini curvatures respectively, $g$ is the metric determinant, $S_m$ is the matter action. Here the Palatini curvature $\mathfrak{R}$ is defined as a function of $g_{\mu\nu}$ and the independent connection $\hat\Gamma^\alpha_{\mu\nu}$:
	\begin{equation}\label{re}	
	\mathfrak{R}=g^{\mu\nu}\mathfrak{R}_{\mu\nu}	=g^{\mu\nu}\bigl(\hat\Gamma^\alpha_{\mu\nu,\alpha}-\hat\Gamma^\alpha_{\mu\alpha,\nu}+\hat\Gamma^\alpha_{\alpha\lambda}\hat\Gamma^\lambda_{\mu\nu}-\hat\Gamma^\alpha_{\mu\lambda}\hat\Gamma^\lambda_{\alpha\nu}\bigr).
	\end{equation}
	
	Like in the pure metric and Palatini cases, the hybrid f(R)-gravity~(\ref{act}) can be rewritten in a scalar-tensor representation (for details see~\cite{hybrid1,hybrid3}):
	\begin{equation}\label{stact1}
	S=\frac{1}{2k^2}\int d^4x\sqrt{-g}\biggl[(1 + \phi)R + \frac{3}{2\phi}\partial_\mu \phi \partial^\mu \phi - V(\phi)\biggr]+S_m,	
	\end{equation}
	where $\phi$ is a scalar field and $V(\phi)$ is a scalar field potential. This potential  $V(\phi)$ is related to the Palatini curvature $\mathfrak{R}$ by the following expression $V_{\phi}=\mathfrak{R}$, where $V_{\phi}$ is the derivative of the potential $V(\phi)$ with respect to the scalar field $\phi$ (for details see~\cite{hybrid1,hybrid3}). 
	
	Then the metric and scalar field equations take the following forms~\cite{hybrid1,hybrid3}:
\begin{wide}
	\begin{eqnarray}
 	&&(1+\phi)R_{\mu\nu}=k^2 \left(T_{\mu\nu}-\frac{1}{2}g_{\mu\nu}T\right)-\frac{3}{2\phi}\partial_\mu\phi\partial_\nu\phi +\frac{1}{2}g_{\mu\nu}  \biggl[V(\phi)+\nabla_\alpha\nabla^\alpha\phi  \biggr]+\nabla_\mu\nabla_\nu\phi,\label{feh}\\
 	&&\nabla_\mu\nabla^\mu\phi-\frac{1}{2\phi}\partial_\mu\phi\partial^\mu\phi-\frac{\phi[2V(\phi)-(1+\phi)V_\phi]}{3}=-\frac{k^2}{3}\phi T_{\rm m},\label{fephi}
	 \end{eqnarray}
	 \end{wide}
	where $T_{\mu\nu}$ and $T$ are the energy-momentum tensor and its trace respectively.
	
	\section{Polarization of gravitational waves}\label{sec3}
	Gravitational radiation propagates by plane waves  in a vacuum. To obtain the linearized field equations in vacuum ($T_{\mu\nu}=0$) we consider the following perturbations of a scalar field and metric tensor:
	\begin{equation}\label{decompos}
	\phi=\phi_0+\varphi,\qquad\ g_{\mu\nu}=\eta_{\mu\nu}+h_{\mu\nu},
	\end{equation}
	where $\phi_0$ is the asymptotic background value of the scalar field far away from a source, $\eta_{\mu\nu}$ is the Minkowski background, $h_{\mu\nu}$ and $\varphi$ are small perturbations of tensor and scalar fields respectively. In the general case, $\phi_0$ is not a constant, but is a function of time $\phi(t)$. However, this dependence can be neglected whenever its characteristic time scale is very long compared with the dynamical time scale associated with the local system itself. Thus, $\phi_0$ is assumed to be constant.
The scalar potential $V(\phi)$ could be expanded in a Taylor series around the background value of the scalar field $\phi_0$ like
	\begin{equation}\label{V}
	V(\phi)=V_0+V'\varphi+\frac{V''\varphi^2}{2!}+\frac{V'''\varphi^3}{3!}...,
	\end{equation}
	hence its derivative with respect to $\varphi$ will be the following:
\begin{equation}
V_\phi=V'+V''\varphi+V'''\varphi^2/2.
\end{equation}
	
	Taking into account the expressions (\ref{decompos}) the linearized field equation for the scalar field~(\ref{fephi}) takes the form:
	\begin{equation}\label{fephi2}
	\left(\nabla^2-m_\varphi^2\right)\varphi=0,
	\end{equation}
		where we denote 
\begin{equation}
m_\varphi^2=[2V_0-V'-(1+\phi_0)\phi_0V'']/3
\end{equation}
 as a scalar field mass.	
		
	The linearized equations for the metric are given by \cite{polarization}
	\begin{equation}\label{feh002}
	\overline{G}_{\mu\nu}=\cfrac{1}{1+\phi_0}(\nabla_\mu\nabla_\nu\varphi-\eta_{\mu\nu}\Box\varphi),
	\end{equation}
where $\overline{G}_{\mu\nu}$ is the linearized Einstein tensor:
\begin{align}\label{gmn}
	\overline{G}_{\mu\nu}&=\overline{R}_{\mu\nu}-\cfrac{1}{2}g_{\mu\nu}\overline{R},\\
\overline{R}_{\mu\nu}&=\cfrac{1}{2}(\partial_\mu\partial_\alpha h^\alpha_\nu+\partial_\nu\partial_\alpha h^\alpha_\mu-\partial_\mu\partial_\nu h-\Box h_{\mu\nu}),\\
\overline{R}&=\partial_\mu\partial_\nu h^{\mu\nu}-\Box h.
	\end{align}
Further we apply the transformations:
\begin{align}\label{theta}
\theta_{\mu\nu}&=h_{\mu\nu}-\cfrac{1}{2}\eta_{\mu\nu}h-\cfrac{1}{1+\phi_0}\eta_{\mu\nu}\varphi,\\
\theta&=-h-4\cfrac{1}{1+\phi_0}\varphi.
\end{align}
The choice of the transverse gauge 
\begin{equation}
\partial_\mu \theta^{\mu\nu}=0
\end{equation}
 reduces the field equation (\ref{feh}) to the following form:
\begin{equation}\label{fet}
\Box\theta_{\mu\nu}=0.
\end{equation}

Equations (\ref{fet}) and (\ref{fephi2}) describe the null  gravitational wave in hybrid metric-Palatini f(R)-gravity. The field $\theta_{\mu\nu}$ is responsible for the familiar massless graviton and tensor part contains two polarization states: the plus and cross
ones. The scalar field $\phi$ is massive, and it decouples from the massless tensor field. Suppose the massless and the massive modes both propagate in the +z direction with the wave vectors
\begin{equation}
k^\mu=(\Omega,0,0,\Omega), \ \ \ \ p^{\mu}=(p_t,0,0,p_z), 
\end{equation}
respectively, where the dispersion relation for the scalar field is 
\begin{equation}
p_t^2-p_z^2=m^2.
\end{equation}
 Then the propagation speed of the massive scalar field is 
\begin{equation}
v=\sqrt{p_t^2-m^2}/p_t. 
\end{equation}
The plane wave solutions of equations (\ref{fet}) and (\ref{fephi2})  take the following forms:
\begin{align}\label{sol}
\theta_{\mu\nu}&=q_{\mu\nu}\exp^{-ik_\alpha x^\alpha},\\
\varphi&=\phi_0\exp^{-ip_\alpha x^\alpha},\ \nonumber
\end{align}
where $\phi_0$ and $q_{\mu\nu}$ are the amplitudes of the waves with 
\begin{equation}
k^\nu q_{\mu\nu}=0, \ \ \ \ \eta^{\mu\nu}q_{\mu\nu}=0.
\end{equation}

The main purpose of this work is to determine the number of polarization states in hybrid f(R)-gravity. These polarization states are contained in the perturbation tensor $h_{ij}$. For a wave traveling in the z-direction, these polarizations become
\begin{equation}
h_{ij} = \left(
\begin{array}{ccc}
h_b+h_+ & h_\times & h_x\\
 h_\times & h_b-h_+& h_y\\
 h_x & h_y & h_L
 \end{array}
 \right),
\end{equation}
where $h_+$ is plus polarization mode (tensor),  $h_\times$ is cross polarization mode (tensor), $h_b$ is transverse breathing polarization mode (scalar), $h_L$ is longitudual polarization mode (scalar). 

 The gravitational waves can influence the distance between the freely moving test particles. This influence is described by the geodesic deviation equation. Assuming that the distance $x^i<<\lambda$, where $\lambda$ is the wavelength of the GWs, and the motion of test particles is slow, the geodesic deviation equation becomes the approximate form 
\begin{equation}
d^2x^i/dt^2 = -R_{0i0j}x^j,
\end{equation}
 where $R_{0i0j}$ is the electric part of Riemann tensor generated by the GWs. Also The GW field $h_{ij}$ is defined by the Riemann tensor, like 
\begin{equation}
\partial^2 h_{ij}/\partial t^2 = -2R_{0i0j}.
\end{equation}
 For further calculations it is necessary to use the linear order of the Riemann tensor
\begin{equation}
\overline{R}_{\mu\nu\alpha\beta}=\cfrac{1}{2}(\partial_\mu\partial_\beta h_{\alpha\nu}+\partial_\nu\partial_\alpha h_{\mu\beta}-\partial_\mu\partial_\alpha h_{\nu\beta}-\partial_\nu\partial_\beta h_{\mu\alpha}).
\end{equation}

Calculation of the electric part of Riemann tensor according to the expressions (\ref{sol}, \ref{theta}) gives the following relations
\begin{wide}
\begin{equation}
R_{0i0j} = \left(
\begin{array}{ccc}
-\frac{1}{2(1+\phi_0)}p_t^2\varphi+\frac{1}{2}\Omega^2\theta_{xx}& \frac{1}{2}\Omega^2\theta_{xy} & 0\\
\frac{1}{2}\Omega^2\theta_{xy} & -\frac{1}{2(1+\phi_0)}p_t^2\varphi-\frac{1}{2}\Omega^2\theta_{xx}& 0\\
0 & 0 &-\frac{1}{2(1+\phi_0)}m^2\varphi
 \end{array}
 \right).
\end{equation}
\end{wide}

To study the polarizations caused by the scalar field, one sets $\theta_{\mu\nu}=0$. In the massless case $(m = 0)$, the component of Riemann tensor $R_{tztz}$ is equal to zero, longitudual mode vanishes, and the scalar field excites only the transverse breathing polarization. In the massive case, it is possible to perform a Lorentz boost such that $p_z = 0$, and work in the rest frame of the scalar field. In this rest frame, $p_t^2=m^2$ and $R_{txtx} = R_{tyty} = R_{tztz} \neq 0$. Moreover, the longitudinal and breathing modes coincide in this case. Thus, we obtain that in the chosen reference frame, the longitudinal mode coincides with the breathing one. Due to the absence of a preferred frame, we can conclude that the scalar mode is a mixture of two modes in other  reference frames, which does not contradict the conclusion about the number of degrees of freedom of the theory.

Let us summarize. In this section, we found polarization states predicted by hybrid f(R)-gravity. For this purpose, first of all we obtained linearized field equations in vacuum and their solutions. All polarization states of a gravitational wave are contained in $h_{ij}$. There is a relation 
\begin{equation}
\partial^2 h_{ij}/\partial t^2 = -2R_{0i0j}
\end{equation}
 between the tensor perturbations of the metric and the electrical part of the Riemann tensor. This connection allowed us to find the form of polarization modes. For these aims we used the definition of the Riemann tensor and the solutions of field equations. All reference frames are equal in hybrid f(R)-gravity, and we passed to the rest frame of the massive scalar field. And in this frame we have established that 
\begin{equation}
p_t^2=m^2,
\end{equation}
 which means 
\begin{equation}
-\frac{1}{2(1+\phi_0)}m^2\varphi=-\frac{1}{ 2(1+\phi_0)}p_t^2\varphi. 
\end{equation}
These expressions in the electrical part of the Riemann tensor are responsible for the scalar longitudinal and transverse breathing  modes, respectively. Since these modes are identical, they are manifestations of the same scalar mode. Thus, we have obtained that two tensor and one scalar degrees of freedom correspond to two tensor and one scalar polarization states. This conclusion removes all contradictions. 

\section{Speed of gravitational waves}\label{sec4}
The GW170817 event gave the possibility to determine the propagation velocity of gravitational waves with unprecedented accuracy \cite{gw2}. As a result, many theories were closed or restricted, including Horndeski model \cite{dead}. Within the framework of this theory, an expression for the propagation velocity of tensor perturbations  was obtained \cite{kase, latosh}. It depends on the general functions of the scalar field included in the structure of Horndeski theory:
\begin{equation}
c_t^2=\cfrac{2G_4-(\dot\phi)^2G_{5,\phi}-(\dot\phi)^2\ddot\phi G_{5,X}}{2G_4-2(\dot\phi)^2G_{4,X}+(\dot\phi)^2G_{5,\phi}-H(\dot\phi)^3G_{5,X}}.
\end{equation}
This expression describes the speed of tensor perturbations $c_t$ propagating over a cosmological background, 
\begin{equation}
X=-1/2\partial_\mu\phi\partial^\mu\phi,
\end{equation}
 $H$ is Hubble constant.

Hybrid f(R)-gravity is a special case of Horndeski theory. The transition functions were obtained earlier \cite{memnras}:
\begin{equation}\label{hybrid} 
	G_2=-\cfrac{3X}{G\phi}-V(\phi),\ G_3=0,\ G_4=\cfrac{1+\phi}{G},\ G_5=0.
	\end{equation} 
	Using these expressions, it is possible to obtain the speed of tensor perturbations in hybrid f(R)-gravity:
	\begin{equation}
c_t^2=\cfrac{2\frac{1+\phi}{G}}{2\frac{1+\phi}{G}}=1.
\end{equation}
Thus, gravitational waves in hybrid f(R)-gravity propagate at the speed of light. The hybrid f(R)-theory is in full agreement existing experimental data.
	
	\section{Conclusions and discussion}\label{sec:conclusions}

	This work is devoted to the issue of determining the number of polarization states predicted by hybrid f(R)-gravity. This theory belongs to the f(R)-family of theories and unites  metric and Palatini approaches. Hybrid f(R)-gravity can be represented as a massive scalar-tensor theory \cite{hybrid1, hybrid3}. Earlier, the question about the discrepancy between the number of polarization modes in massive scalar-tensor theories and degrees of freedom was raised \cite{hougong}. This issue has been studied in detail on the example of Horndeski theory \cite{hougong}. It was shown that the theory has two tensor  and one scalar degrees of freedom. The scalar degree of freedom is responsible for the mixture of transverse breathing and longitudinal scalar modes. In addition, it was shown in the work \cite{hougong} that the Eardley classification is not applicable to scalar-tensor theories with a massive scalar field.
	
For hybrid  f(R)-gravity, the number of polarization states was previously obtained \cite{polarization}. It was concluded,  there are transverse breathing and longitudinal modes in addition to tensor cross and plus modes. However, it was determined that hybrid f(R)-gravity has three degrees of freedom \cite{koivisto1}. Therefore, the main purpose of this article was to explain the existing discrepancy.

All polarization modes predicted by the theory are contained in the electrical part of the Riemann tensor. After carrying out the corresponding calculations, we found that both longitudinal and transverse breathing scalar modes will be present in the hybrid f(R)-gravity. However, in the rest frame of a massive scalar field, two scalar polarization states become identical. Thus, we can conclude that both scalar modes are manifestations of the same scalar degree of freedom and represent a mixture of polarization states of a gravitational wave. Therefore, there is no contradiction between the number of degrees of freedom and the number of polarization states in hybrid f(R)-gravity.

In conclusion, we can compare the results obtained for the hybrid f(R)-gravity with the results obtained for the metric and Palatini f(R)-theories. Within the framework of metric f(R)-gravity, it was shown that there are longitudinal and transverse breathing scalar modes, in addition to the tensor plus and cross modes. These modes are identical in all reference frames, since there is the relationship between the mass of the scalar field in the scalar-tensor representation of this theory and the model parameter $\alpha$: $m^2=-1/6\alpha$ \cite{polarmetric}. In addition, in metric f(R)-gravity, the question of the propagation of primary gravitational waves was considered in detail. It was shown that for such models the energy spectrum of primordial gravitational waves was slightly expanded compared to the GR spectrum, but the predicted spectrum is much lower than the lowest sensitivity values of future experiments on the detection of gravitational waves at high frequencies \cite{odintsovpolar}. In the Palatini formalism, there are only tensor plus and cross modes, and such models are indistinguishable from GR \cite{palatinipolar}.

One of the most effective tests of modified theories of gravity  is the calculation of the GWs speed. This test made it possible to cut off a large number of theories, including the scalar-tensor ones. Moreover, some authors argue that only the Brans-Dicke theory with a massive scalar field is able to pass this test \cite{dead}. In this paper, we calculated the propagation velocity of tensor perturbations in hybrid f(R)-gravity and showed that the velocity is identically equal to the speed of light. Thus, hybrid f(R)-gravity is in full agreement with the available experimental data.

This work is the first step in the study of gravitational-wave radiation in the framework of hybrid f(R)-gravity. For a realistic test of the theory, it is necessary to obtain gravitational wave templates and compare them with the available LIGO and VIRGO observational data. Such investigation is supposed to become an important step in development of a new powerful formalism for testing other alternative theories of gravity in the strong field regime of coalescing binary systems.

The author thanks N. A. Avdeev and B.N.Latosh for discussions and comments on the topics of this paper. The work was supported by the Foundation for the Advancement of Theoretical Physics and Mathematics ``BASIS''.


\begin{thebibliography}{99}
\bibitem{gw}
B.P.~Abbott et al.,  Phys. Rev. Lett. {\bf 116(6)}, 061102 (2016).
\bibitem{gw1}
B.P. Abbott et al.,  Phys. Rev. Lett. {\bf 119(14)}, 141101 (2017).
\bibitem{liu}
T. Liu, W. Zhao, Y. Wang, Phys. Rev. {\bf D 102}, 124035 (2020).
\bibitem{eardley}
D.M. Eardley, D.L. Lee, A.P. Lightman,  Phys. Rev. {\bf D 8}, 3308 (1973).
\bibitem{eardley1}
D.M. Eardley, D.L. Lee, A.P. Lightman, R.V. Wagoner, C.M. Will,  Phys. Rev. Lett. {\bf 30}, 884 (1973).
\bibitem{maggiore}
M. Maggiore and A. Nicolis, Phys. Rev. {\bf D 62}, 024004 (2000).
\bibitem{hougong}
S. Hou, Y. Gong, Y. Liu, Eur. Phys. J. {\bf C 78},  378 (2018).
\bibitem{dong}
Y.-Q. Dong, Y.-X. Liu, 	Phys.Rev. {\bf D 105},  6, 064035 (2022).
\bibitem{horndeski}
G.W. Horndeski, Int. J. Theor. Phys. {\bf 10}, 363 (1974).
\bibitem{hybrid1}
		T.~Harko, T.~S.~Koivisto, F.~S.~N.~Lobo and G.~J.~Olmo,  Phys. Rev. {\bf D 85}, 084016 (2012).
		\bibitem{hybrid2}
		T. Harko, F.S. N. Lobo, Int. J. Mod. Phys. {\bf D 29}, 2030008 (2020).
		 \bibitem{fr1}
		P.~G.~Bergmann, International Journal of Theoretical Physics {\bf 1},  25 (1968).
		
		\bibitem{fr2}
		A.~De Felice and S.~Tsujikawa, Living Reviews in Relativity {\bf 13},  3 (2010).
		\bibitem{fr3}
		S.~Nojiri, S.~D.~Odintsov and V.~K.~Oikonomou,	Phys.Rept. {\bf692},  1-104 (2017).
\bibitem{starobinsky}
		A.~A.~Starobinsky, Phys. Lett. B {\bf 91},  99 (1980).
		\bibitem{odintsov1}
		S.~Nojiri and S.~D.~Odintsov, Phys. Rev. D {\bf 68}, 123512 (2003).
		
		\bibitem{odintsov2}
		  F.~Briscese, E.~Elizalde, S.~Nojiri and S.~D.~Odintsov, Phys. Lett. B {\bf 646}, 105 (2007).
		  
		 
		   
		   \bibitem{odintsov3}
		   S.~Nojiri and S.~D.~Odintsov, Phys. Lett. B {\bf 657}, 238 (2007).
		   
		   \bibitem{odintsov4}
		   S.~Nojiri and S.~D.~Odintsov, Phys. Rev. D {\bf 77}, 026007 (2008). 
		   
		   
		   \bibitem{odintsov5}
		    S.~Nojiri, S.~D.~Odintsov and D.~Saez-Gomez, Phys. Lett. B {\bf 681}, 74 (2009).
		    
		  \bibitem{odintsov6}
		   G.~Cognola, E.~Elizalde, S.~D.~Odintsov, P.~Tretyakov and S.~Zerbini, Phys. Rev. D {\bf 79}, 044001 (2009).
		   
		   \bibitem{odintsov7}
		   G.~Cognola, E.~Elizalde, S.~Nojiri, S.~D.~Odintsov, L.~Sebastiani and S.~Zerbini, Phys. Rev. D {\bf 77}, 046009 (2008). 
		   	    \bibitem{saez}
		   D.~Saez-Gomez, Gen. Rel. Grav. {\bf 41}, 1527 (2009). 
		   \bibitem{odintsov8}
		   S.~D.~Odintsov, D.~Saez-Gomez and G.~S.~Sharov,  Eur.Phys.J. C {\bf 77}, 862 (2017).
\bibitem{chiba}
		T.~Chiba, Phys. Lett. B {\bf 575},  1 (2003).	 
		 \bibitem{khoury}
		  J.~Khoury and A.~Weltman, Phys. Rev. Lett. {\bf 93}, 171104 (2004).
		     \bibitem{khoury1}
		J.~Khoury and A.~Weltman, Phys. Rev. D {\bf69}, 044026 (2004).
		\bibitem{hu}
		 W.~Hu and I.~Sawicki, Phys. Rev. D {\bf76}, 064004 (2007).
\bibitem{koivisto}
		T.~Koivisto and H.~Kurki-Suonio,  Class. Quantum Grav. {\bf 23}, 2355 (2005).
				 \bibitem{hybrid3}
		S.~Capozziello et al., Hybrid metric-Palatini gravity, Universe {\bf 1}, 199 (2015).
		\bibitem{hybrid4}
		S. Capozziello, T. Harko, T. S. Koivisto, F. S. N. Lobo and G. J. Olmo, JCAP{\bf  07}, 024 (2013).
		\bibitem{hybrid5}
		S. Capozziello, T. Harko, T. S. Koivisto, F. S. N. Lobo and G. J. Olmo, Astroparticle Physics {\bf 50-52C}, 65 (2013).
		\bibitem{hybrid6}
		S. Capozziello, T. Harko, T. S. Koivisto, F. S. N. Lobo and G. J. Olmo, Phys. Rev. {\bf D 86}, 127504 (2012).
		\bibitem{bronnikov}
		K. A. Bronnikov, S. V. Bolokhov and M. V. Skvortsova, Grav. Cosmol. {\bf 26}, 212-227 (2020).
		\bibitem{danila}
		B. Danila, T. Harko, F. S. N. Lobo and M. K. Mak, Phys. Rev. {\bf D 95}, 044031 (2017).
		\bibitem{meppn}
		P. I. Dyadina, S. P. Labazova, S. O. Alexeyev, JETP {\bf156}, 5, 905-917 (2019).
		\bibitem{meshapiro}
		 P. I. Dyadina, S. P. Labazova,  JCAP {\bf01}, 029 (2022).
		 \bibitem{memnras}
		 P. I. Dyadina, N. A. Avdeev and S. O. Alexeyev, MNRAS {\bf483}, 947 (2019).
		 \bibitem{meppk}
		 N. Avdeev, P. Dyadina, S. Labazova, JETP {\bf158}, 4, 613-625 (2020).
		 \bibitem{polarization}
H. R. Kausar,  Astrophys. Space Sci. {\bf363}, 11, 238 (2018).
\bibitem{koivisto1}
T. S. Koivisto, N. Tamanini, Phys. Rev. {\bf D 87}, 104030 (2013).
\bibitem{gw2}
B.P. Abbott et al., Phys. Rev. Lett. {\bf119(16)}, 161101 (2017).

\bibitem{dead}
J. M. Ezquiaga, M.Zumalacarregui, Phys. Rev. Lett. {\bf119}, 251304 (2017).
\bibitem{kase}
R. Kase, S. Tsujikawa, Int. J. Mod. Phys. {\bf D28}, 05, 1942005 (2019).
\bibitem{latosh}
B. Latosh, Eur. Phys. J. {\bf C 80}, 845 (2020).
\bibitem{polarmetric}
D. Liang, Y. Gong, S. Hou, Y. Liu, Phys. Rev. {\bf D 95}, 104034 (2017).
\bibitem{odintsovpolar}
S.D. Odintsov, V.K. Oikonomou, F.P. Fronimos, Physics of the Dark Universe {\bf 35}, 100950 (2022).
\bibitem{palatinipolar}
M.E.S. Alves, O. D. Miranda, J. C.N. de Araujo, Physics Letters B {\bf 679},  4,  401 (2009).
		
		

	
\end{thebibliography}
\end{document}